\documentclass[a4paper]{jpconf}
\usepackage{graphicx}
\usepackage{cite}
\usepackage{hyperref}
\begin{document}
\title{Prospects of nuclear clustering studies via dissociation of relativistic nuclei in nuclear track emulsion}

\author{D A Artemenkov$^1$,V Bradnova$^1$, E Firu$^2$, M Haiduc$^2$, \mbox{N K Kornegrutsa$^1$,} A I Malakhov$^{1}$, E Mitsova$^{1,3}$, A Neagu$^2$, \mbox{N G Peresadko$^4$,} V V Rusakova$^1$, R Stanoeva$^{3,5}$, A A Zaitsev$^{1,4}$, \mbox{I G Zarubina$^1$} and P I Zarubin$^{1,4}$ }

\address{$^1$Joint Institute for Nuclear Research, Dubna, Russia}
\address{$^2$Institute of Space Sciences, Magurele, Romania}
\address{$^3$Southwestern University, Blagoevgrad, Bulgaria}
\address{$^4$P.N. Lebedev Physical Institute, Moscow, Russia}
\address{$^5$Institute for Nuclear Research and Nuclear Energy, Sofia, Bulgaria}

\ead{zarubin@lhe.jinr.ru}

\begin{abstract}
Status and prospects of nuclear clustering studies by dissociation of relativistic nuclei in nuclear track emulsion are presented. The unstable $^{8}$Be and $^{9}$B nuclei are identified in dissociation of the isotopes $^{9}$Be, $^{10}$B, $^{10}$C and $^{11}$C, and the Hoyle state in the cases $^{12}$C and $^{16}$O. On this ground searching for the Hoyle state and more complex $\alpha$-particle states in the dissociation of the heavier nuclei is suggested. A detailed study of a low-density baryonic matter arising in dissociation of the heaviest nuclei is forthcoming long-term problem. An analysis of nuclear fragmentation induced by relativistic muons is proposed to examine the mechanism dissociation. 
\end{abstract}

\section{Introduction}
Theories of the matter arising due to nucleon clustering in the lightest nuclei under conditions of a lowest baryonic density \cite{1} and $\alpha$-particle Bose–Einstein condensate \cite{2} have been developed. Such states may be necessary stages toward synthesis of heaviest nuclei. Thereby, a studying of lowest energy nucleon-cluster states including unbound nuclei becomes a challenge for the experiment. Relevant ensembles can be studied in dissociation of relativistic nuclei in a nuclear track emulsion (NTE) (review \cite{3}). Necessary measurements of emission angles of relativistic fragments H and He and their identification by the multiple scattering are accessible in NTE only. Perhaps, a nuclear matter similar in thermodynamics and isotopic composition with a supernova can be re-created in dissociation of heaviest nuclei. The hypothesis about reproduction of microscopic pieces of such a matter in dissociation of relativistic nuclei in NTE is in a focus of the BECQUEREL experiment in JINR. Next, the findings and suggestions will be highlighted.

\section{Development of experiment}
The \href{http://becquerel.jinr.ru/}{BECQUEREL} experiment is based on the NTE technique created in the 1950s and not lost till now. Exposures of NTE in beams of relativistic nuclei started in the 70s at the JINR Synchrophasotron and LBL Bevalac and in the 90s at BNL AGS and CERN SPS. Most destructive collisions as meeting the highest concentration of matter and energy attracted special attention. The subsequent development in this direction is widely known. At the same time, results obtained in the 70–90s as well as the exposed NTE layers retain uniqueness in the respect of projectile dissociation. 

The projectile nucleus structure is manifested holistically in peripheral collisions. Using NTE layers exposed at the JINR Nuclotron the features $^{7,9}$Be, $^{8,10,11}$B, $^{10,11}$C and $^{12,14}$N manifested in probabilities of dissociation channels were studied \cite{3,4}. Events of such a type are observed as often and fully as violent collisions. The most peripheral ones called coherent dissociation or ``white stars'' are accompanied neither by target nuclei fragments nor produced mesons.

Calculation of the invariant mass of an ensemble of relativistic fragments allows one to set its inner energy scales in it. In general, this variable is defined as the sum of all products of 4-momenta $P_{i,k}$ of the fragments $M^{*2}$ = $\Sigma(P_i \cdot P_k)$. It is convenient to subtract the initial nucleus mass or fragment mass sum $Q = M^* - M$. In the projectile fragmentation cone the components $P_{i,k}$ can be determined by assuming conservation of a primary nucleus speed (or momentum per nucleon) by its fragments. Reconstruction of the invariant mass of relativistic decays of the unstable nuclei $^{8}$Be and $^{9}$B confirms this approximation and allows establishing their contribution to the dissociation pattern. On the basis of decays $^{9}$B $\to$ $p^{8}$Be identified in the $^{10}$C dissociation the $^8$Be criteria $Q_{2\alpha}$ $<$ 200 keV is confirmed. Selected under the cleanest conditions it takes into account practical resolution and accepted approximations, including the kinematic ellipse of $^{8}$Be decay.

The conclusion is that the unstable nuclei do contribute in the nuclear structure. These observations suggest the possibility of expanding the scenarios of light isotope synthesis involving the unstable states. In nucleosynthesis chains, they may serve necessary ``transfer stations'', the passage through which gets ``imprinted'' in the nuclei formed. Until now, the unstable ``transfer station'' is recognized only in the famous chain 3$\alpha$ $\to$ $\alpha^{8}$Be $\to$ (Hoyle state) $\to$ $^{12}$C.

\section{Hoyle state in dissociation of light nuclei}
The successful reconstruction of the $^8$Be and $^9$B decays allows one to take the next step — to search in relativistic dissociation $^{12}C$ $\to$ 3$\alpha$ in NTE for triples of $\alpha$-particles in the Hoyle state (HS) \cite{5,6,7}. This state is the second and first unbound excitation 0$_2^+$ of the $^{12}$C nucleus. A research status of the Hoyle state (HS) is presented in the review \cite{8}. The HS features such as isolation in the initial part of the $^{12}$C excitation spectrum, extremely small values of decay energy and width (378 keV and 8.5 eV) indicate its similarity to $^{8}$Be (91 keV and 5.6 eV). In relativistic case the smallest energy values of the $^{8}$Be and HS decays are projected to narrowest flying $\alpha$-particle pairs and triples, respectively. The $^{8}$Be nucleus is an indispensable product of the HS decay. The values of the $^{8}$Be and HS widths (or their lifetimes) indicate the pronounced sequential decay of $^{12}$C(0$_2^+$) $\to$ $^{8}$Be $\to$ 2$\alpha$. Identification of $^{8}$Be can serve for the HS candidate pre-selection.

It can be assumed that HS is not limited to $^{12}$C excitation but it can also appear as a 3$\alpha$-particle analogue of $^{8}$Be in relativistic fragmentation of heavier nuclei. The current interest is motivated by the concept of $\alpha$-particle Bose-Einstein condensate (review \cite{2,9}) of such a condensate. The $^{8}$Be nucleus, HS and, then, the $^{16}$O  0$_6^+$ excitation 700 keV above the 4$\alpha$ threshold are considered as its simplest forms. The condensate can decompose via the decay sequence {$^{16}$O(0$_6^+$) $\to$ $^{12}$C(0$_2^+$) $\to$ $^{8}$Be $\to$ 2$\alpha$. Then, identification of $^{12}$C(0$_2^+$) can be applied to tag candidates $^{16}$O ($0_6^+$). In such a way, this idea can be extended.

The HS generation may reflect both the presence of the 3$\alpha$-particle-like 0S-state in the parent nucleus as well as arise through the excited fragment $^{12}$C$^*$($\to$3$\alpha$). The next question is as follows. Can the fragmentation of relativistic nuclei serve as a ``factory'' for the generation of ensembles of $\alpha$-particles of increasing multiplicity at the lower limit of nuclear temperature? Based on data obtained on dissociation of $^{12}$C \cite{10} and $^{16}$O \cite{11} and their complement in the case of $^{12}$C a relevant analysis was done. Its summary as follows.

Determining the invariant mass of the $\alpha$-particle triples $Q_{3\alpha}$ is sufficiently precise to identify HS against the higher $^{12}$C excitations (Fig.\ref{fig.1}). The HS decays contribution to $^{12}$C $\to$ 3$\alpha$ coherent dissociation within $Q_{3\alpha}$ $<$ 0.7 MeV is 12 $\pm$ 2\%. In the case of the coherent dissociation of \mbox{$^{16}$O $\to$ 4$\alpha$} it reaches 22 $\pm$ 2\% while the fraction of events $^{16}$O $\to$ 2$^{8}$Be is just 5 $\pm$ 1\%. Thereby, the hypothesis about HS as a universal object similar to $^{8}$Be gets the support. Moreover, rise of $\alpha$-particle combinations $^{16}$O $\to$ 4$\alpha$ leads the enhanced HS contribution. An analysis of the invariant masses of $\alpha$-quartets $Q_{4\alpha}$ gives an estimate of the contribution of the state $^{16}O$ $0_6^+$ equal to 7 $\pm$ 2\%. Hence, the direct dissociation $\alpha$ + HS dominates. Analysis of fragmentation of the $^{22}$Ne nucleus revealed the HS formation only in the 4$\alpha$ channel at 15 $\pm$ 4\%. 

\begin{figure}
	\begin{center}
		\includegraphics*[width=0.65\linewidth]{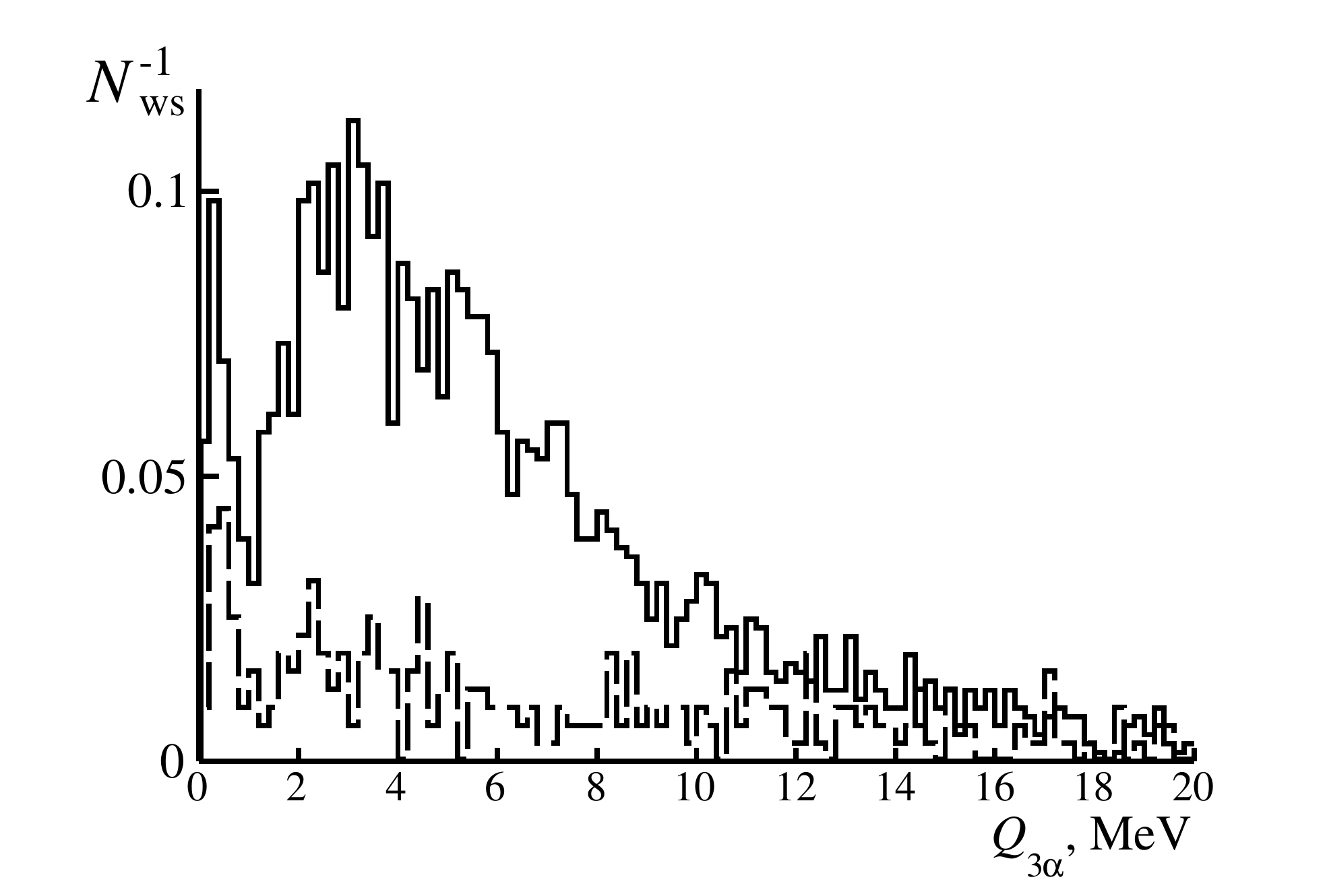}
	\end{center}
\caption{\label{fig.1}Distribution of all $\alpha$-triples over invariant mass $Q_{3\alpha}$ in coherent dissociation \mbox{$^{12}$C $\to$ 3$\alpha$} (314 events, dashed) and $^{16}$O $\to$ 4$\alpha$ (641 events, solid) at 3.65 $A$ GeV; distributions are normalized per number of ``white” stars''.}
\end{figure}

The HS observations inspire testing its universality and searching for heavier $\alpha$-condensate states using available NTE layers exposed to the neighboring nuclei. The closest goal is the $^{14}$N dissociation in which the channel 3He + H leads with the $^{8}$Be contribution of 25\% \cite{12}. The re-analysis is resumed recently in a context of the HS problem and the $^{9}$B contribution. A similar analysis will be carried out in the NTE layers which exposed to relativistic nuclei $^{22}$Ne, $^{24}$Mg and $^{28}$Si. Besides, there is sufficient amount of the NTE layers exposed at CERN SPS to $^{32}$S nuclei at 200 $A$ GeV. Potentially, the solution to the question of the universality of HS will open the horizon of a search for even more complex systems with participation $^{8}$Be and HS. Earlier, the $^{8}$Be formation was identified in exposures to Pb nuclei at the CERN SPS. Then, the challenge is to identify HS.

\section{Composition of dissociation of heavy nuclei}
The studies of lighter nuclei pave a path toward more and more complex ensembles He – H – $n$ produced in dissociation of heavier nuclei. As a limit, dissociation of relativistic nuclei Au, Pb and even U was reliably observed in NTE down to the lightest nuclei and nucleons even without visible excitation of target nuclei. It is possible that this phenomenon confirms the essential role of the long-range quantum electrodynamics interaction. Charges of heavy nuclei make possible multiple photon exchanges and transitions in many-particle state. An alternative scenario of coherent dissociation consists in virtual meson exchanges. Interference of electromagnetic and strong interaction is possible.

Dissociation of heavy nuclei leads to appearance of many-particle states with kinematic characteristics that are of nuclear astrophysical interest and which cannot be formed in other laboratory conditions. Inversion of the ``time arrow'' in such events suggests element synthesizing through the nucleon and lightest nucleus phase. Following to the emission angles the fragment energy in the parent nucleus system covers the temperature range 10$^{8-10}$ K, i. e. from the red giant phase to supernova. The Coulomb repulsion is radically weakened in such rarefied ensembles. Being considered in a macroscopic scale, such a ``nuclear packaging'' can serve as a source of gravitational waves.

It is proposed to use the unique and, together with that, well-proven capabilities of the NTE method for an in-depth study of the peripheral dissociation of heavy nuclei with energy of several GeV per nucleon. At the initial stage, the available 500 $\mu$m thick NTE layers are analyzed using statistics of dozens of peripheral interactions of Kr (2 $A$ GeV, GSI), Au (10 $A$ GeV, BNL) and Pb (159 $A$ GeV, CERN) nuclei to determine the dependence of the neutron contribution on the degree dissociation of these nuclei. To characterize the emerging state, the ratios of relativistic isotopes of $^{1,2,3}$H and $^{3,4}$He have to be determined. While being possible, the time-consuming analysis of the isotopic composition of relativistic fragments by the scattering method was not used in the 90s with pioneering irradiation with Au nuclei at 10.7 $A$ GeV \cite{13,14,15,16}. Besides, although the formation of secondary stars by neutrons in the cone of fragmentation was observed their research tasks were not set. 

In the relativistic dissociation of heavy nuclei, light fragments are formed with a higher charge-to-mass ratio than that of the primary nucleus causing the appearance of associated neutrons. The neutron average range in NTE is 32 cm. These neutrons must detect themselves in the fragmentation cone by secondary stars that do not contain an incoming track. The frequency of such ``neutron'' stars should increase with an increase of a lightest nuclei multiplicity in the fragmentation cone. Reaching dozens, the multiplicity of neutrons in an event can be estimated by a proportional decrease in the mean path to the formation of ``neutron'' stars at lengths of the order of several centimeters. The coordinates of the interaction vertex are determined with an accuracy characteristic of NTE (not worse than 0.5 $\mu$m), which allows the angles of neutron emission to be recovered with the best accuracy Measurements of adjacent tracks can be used to compensate for possible distortion. In the case of complete dissociation of a heavy nucleus, the number of neutrons can be estimated by the isotopic composition of the relativistic fragments H and He. Is the yield of deuterons and tritons binding neutrons significant? The answer to this question may also have practical significance.

\begin{figure}
	\begin{minipage}[h]{0.5\linewidth}
		\includegraphics[width=1\linewidth]{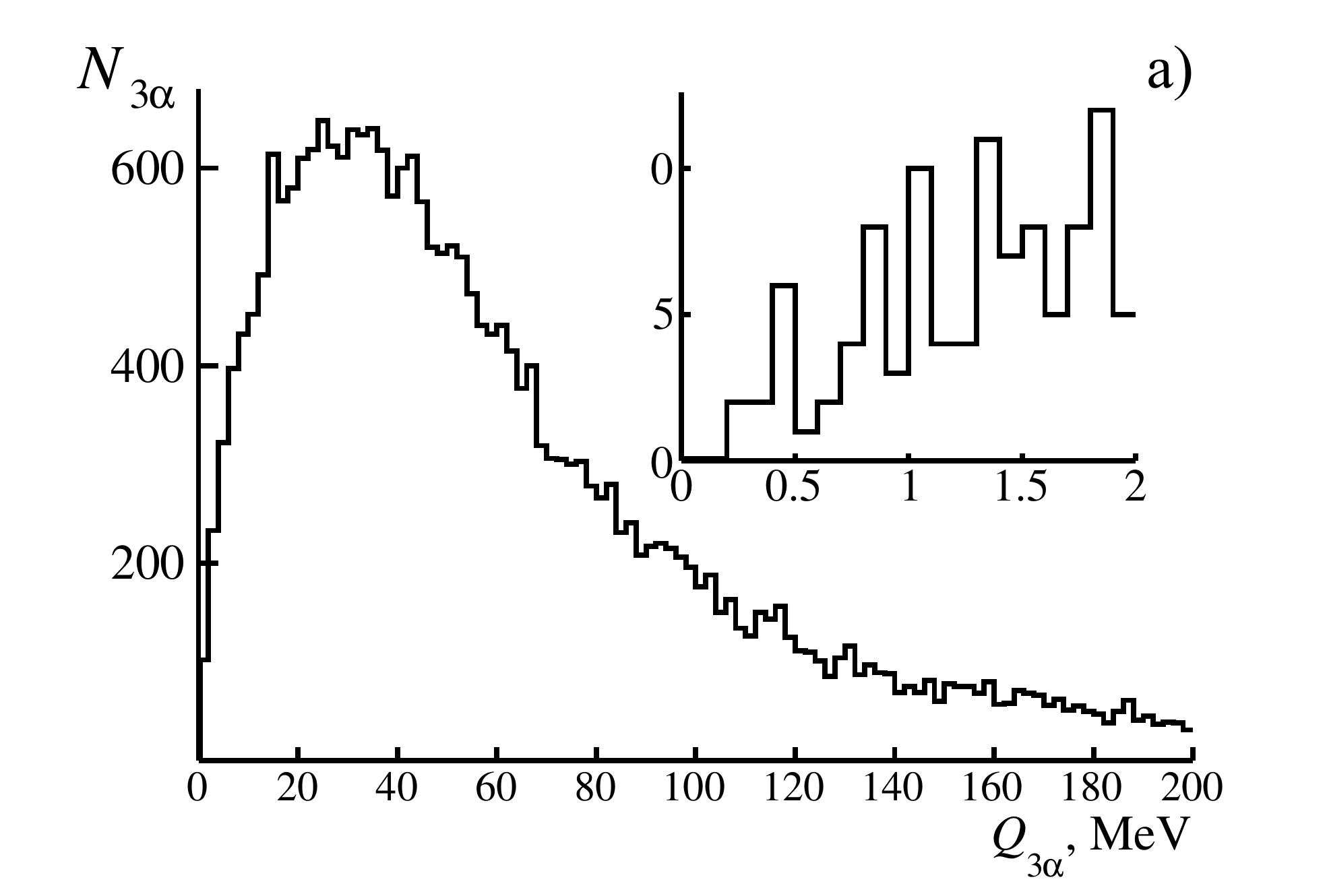}
	\end{minipage}
	\hfill
	\begin{minipage}[h]{0.5\linewidth}
		\includegraphics[width=1\linewidth]{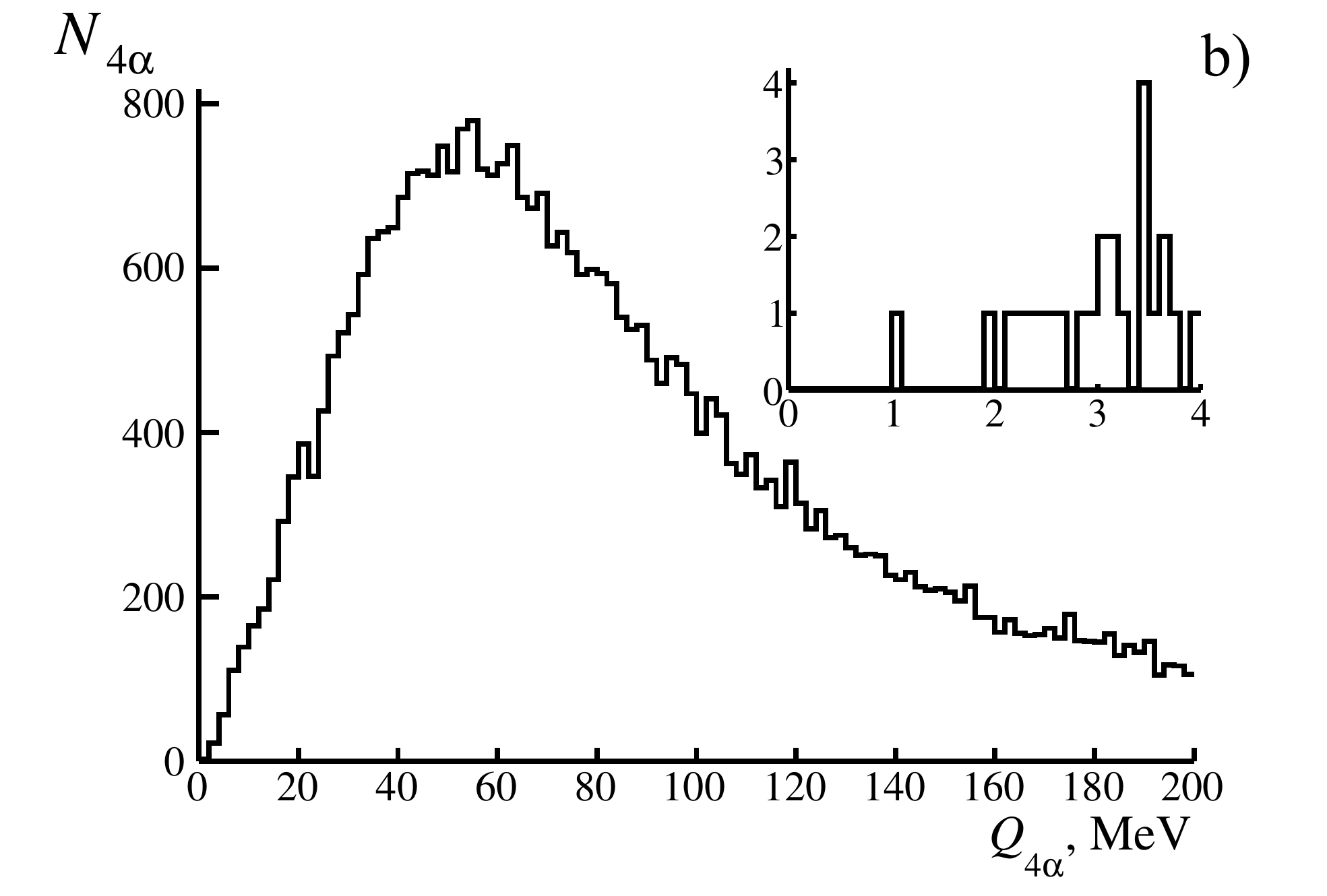}
	\end{minipage}
	\vfill
	\begin{minipage}[h]{1\linewidth}
		\begin{center}
			\includegraphics[width=0.5\linewidth]{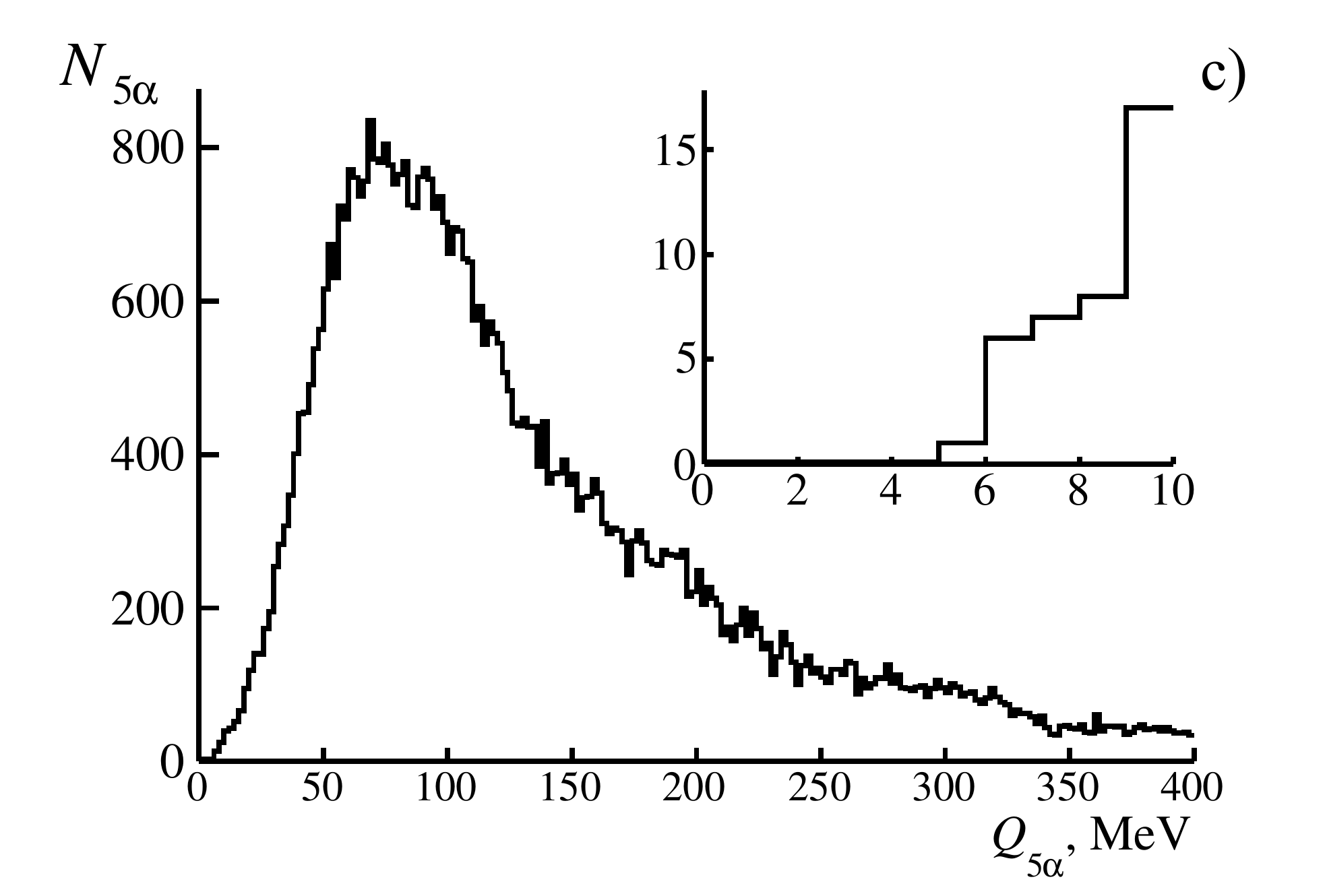}
    \end{center}	
\end{minipage}
	\caption{Distributions over invariant mass $Q$ of $\alpha$-particle triples ({\it a}), fours ({\it b}) and fives ({\it c}) produced in dissociation of 10.7 $A$ GeV Au nuclei.}
	\label{fig.2}
\end{figure}

Angular measurements of the tracks in 1316 interactions of Au nuclei at 10.7 $A$ GeV including 30 ``white'' stars are available for preliminary analysis. This sample includes 843 events with groups of 3 to 16 relativistic $\alpha$ particles in the fragmentation cone. The universality of the invariant mass variable $Q$ allows a uniform analysis of $\alpha$-particle ensembles of increasing complexity. Figure 6 shows the $Q$ distributions for all possible combinations of triples, quartets, and quintets of $\alpha$ particles, including the enlarged regions of small $Q$ values. Of primary interest is the identification of HS decays by the $Q_{3\alpha}$ distribution of all possible combinations of $\alpha$-particle triples.

According to Figure \ref{fig.2}(a) 12 events contains one and only one $\alpha$-triple satisfying the condition $Q_{3\alpha}$ $<$ 0.7 MeV. For these $\alpha$ triples, the average value is $\langle Q_{3\alpha} \rangle$ = (435 $\pm$ 29) keV at RMS 106 keV which does not contradict the HS decay hypothesis. The distribution of these events over the $\alpha$-particle multiplicity $N_{\alpha}$ is as follows: $N_{\alpha}$ = \mbox{4 (2)}, \mbox{6 (2)}, \mbox{7 (2)}, \mbox{8 (2)}, \mbox{9 (1)}, \mbox{11 (1)}, \mbox{16 (1)}. It turns out to be fairly uniform. Then, the isolated $\alpha$-quadruple (Figure \ref{fig.2}(b)) with $Q_{4\alpha}$ = 1 MeV which includes one of the $\alpha$-triples $Q_{3\alpha}$ $<$ 0.7 MeV is identified. As one would expect in the distribution over $Q_{5\alpha}$ (Figure \ref{fig.2}(c)), $\alpha$-fives are not observed up to $Q_{5\alpha}$ $<$ 5 MeV. As a positive fact, one can also note the absence of a combinatorial background.

Thus, the idea of the possibility of HS creation in the dissociation of heavy nuclei is supported and there is the prospect of searching for its heavier analogues. In this regard, it is necessary to increase the statistics of $N_{\alpha}$$-$jets by means of accelerated viewing and to concentrate on angular measurements corresponding to the region of small $Q_{3\alpha}$.

\section{Nuclear fragmentation induced by muons}
The mechanism of dissociation of relativistic nuclei in peripheral interactions remains unclear. It is possible that there is a multiple photon exchange between the nuclei of the beam and the target. The alternative is to exchange virtual mesons. As a critical test, fragmentation of nuclei of the NTE composition under the action of relativistic muons can serve \cite{17,18,19}. In this case, fragmentation may occur as a result of the transition of exchange photons into pairs of virtual mesons. This combination provides long-range action at effective destruction of nuclei and can be extended to peripheral interactions of relativistic nuclei. In this regard, it is necessary to carry out a search for the fullest possible destruction of heavy nuclei of the NTE composition (Ag and Br) under the action of relativistic muons.

\begin{figure}
	\begin{center}
		\includegraphics*[width=0.5\linewidth]{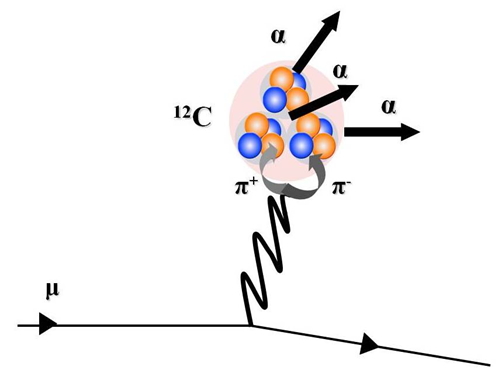}
	\end{center}
	\caption{\label{fig.3} Break-up of $^{12}$C nucleus into three $\alpha$-particles by relativistic muon.}
\end{figure}

Cases of fragmentation of target nuclei into three $b$-particles are most probable for the breakup of $^{12}$C $\to$ 3$\alpha$ (Fig.\ref{fig.3}). In these events, the ranges and angles of emission of $\alpha$-particles are determined on the basis of coordinate measurements of tracks. The $\alpha$-particle energy values are extracted from spline-interpolation of the energy-range calculation using the well-known SRIM model. On this basis, one can obtain the distributions over the invariant mass as well as over the total momentum of pairs and triples of $\alpha$-particles. The procedure of reconstruction of the invariant mass was tested in the reconstruction of the ground state $^{8}$Be$^+_0$ in the NTE exposed to 14.1 MeV neutrons and the first excited $^{8}$Be$^+_2$ in the $^{8}$He implantation in NTE \cite{20}.

The transverse irradiations of NTE layers in the halo of the 160 GeV muon beam with a duration of up to a day were performed at CERN in 2017. However, a muon flux and hadron admixture estimates weren’t provided. In the muon beam of the COMPASS experiment, the hadron admixture does not exceed 10$^{–6}$, and a short exposure of acceptable density in the defocused beam can be suggested. It has been preliminarily established that the distribution over a total transverse momentum of $\alpha$-particle triples produced by the splitting of $^{12}$C nuclei corresponds not to electromagnetic, but to nuclear diffraction. The determination of the 3$\alpha-$splitting cross section is of importance for geophysics, since it will allow testing the hypothesis of the generation of helium in the depths of the Earth’s crust by space muons.

\end{document}